# Model Predictive Control Paradigms for Fish Growth Reference Tracking in Precision Aquaculture*

Abderrazak Chahid[a], Ibrahima N'Doye[a], John E. Majoris[b], Michael L. Berumen[b] and Taous Meriem Laleg-Kirati[a]

[a]Computer, Electrical and Mathematical Sciences and Engineering Division (CEMSE), King Abdullah University of Science and Technology (KAUST), Thuwal 23955-6900, Saudi Arabia.

[b]Red Sea Research Center, Biological and Environmental Science and Engineering Division, King Abdullah University of Science and Technology (KAUST), Thuwal 23955-6900, Saudi Arabia.

ABSTRACT

In precision aquaculture, the primary goal is to maximize biomass production while minimizing production costs. This objective can be achieved by optimizing factors that have a strong influence on fish growth, such as the feeding rate, temperature, and dissolved oxygen. This paper provides a comparative study of three model predictive control (MPC) strategies for fish growth reference tracking under a representative bioenergetic growth model in precision aquaculture. We propose to evaluate three candidate MPC formulations for fish growth reference tracking based on the receding horizon. The first MPC formulation tracks a desired fish growth trajectory while penalizing the feed ration, temperature, and dissolved oxygen. The second MPC optimization strategy directly penalizes the feed conversion ratio (FCR), which is the ratio between food quantity and fish weight gain while minimizing the actual growth state's deviation from the given reference growth trajectory. The third MPC formulation includes a tradeoff between the growth rate trajectory tracking, the dynamic energy and the cost of food. Numerical simulations that integrate a realistic bioenergetic fish growth model of Nile tilapia (*Oreochromis niloticus*) are illustrated to examine the comparative performance of the three proposed optimal control strategies.

## 1. Introduction

Aquaculture is one of the largest and fastest-growing food production sectors in the world and is likely to become the primary source of seafood in the future [4]. As commercial fish production continues to increase, both its impact and reliance on protein sources provided by ocean fisheries are likely to expand. To mitigate these impacts, adequate growth models are relevant for efficient management of aquaculture, as they provide an optimized protocol for feeding and monitoring fish welfare throughout the grow-out cycle from stocking through harvesting [21]. Thus, there is a pressing need to develop precision aquaculture techniques that improve fish farming efficiency by optimizing feeding protocols [17].

Modern aquaculture systems can benefit from the integration of emerging technologies and theory from multiple research disciplines such as marine science and optimal control systems. In the integration of control systems, classical feedback approaches are not directly convenient to most feeding regimes due to the scheduled nature of the feed ration and biological constraint of the aquaculture environment. Hence, the control problem is generally derived as an optimization problem targeting the desired growth trajectory subject to some constraints such as food quantity, environmental parameters, and economic factors, as illustrated in Fig. 1. The integration of new technology-based solutions and policies may help to promote sustainable aquaculture production. Currently, there are no examples of closed-loop precision fish farming systems, which include the different components of observing the fish for decision making [6]. Therefore, the aquaculture industry and researchers aspire to develop strategies that optimize biomass production by monitoring and controlling factors that influence fish growth.

*This work has been supported by the King Abdullah University of Science and Technology (KAUST), Base Research Fund (BAS/1/1627-01-01) to Taous Meriem Laleg and Base Research fund KAUST – AI Initiative Fund.

*Corresponding author: T. M. Laleg-Kirati

abderrazak.chahid@kaust.edu.sa (A. Chahid); ibrahima.ndoye@kaust.edu.sa (I. N'Doye); john.majoris@kaust.edu.sa (J.E. Majoris); michael.berumen@kaust.edu.sa (M.L. Berumen); taousmeriem.laleg@kaust.edu.sa (T. Laleg-Kirati)

ORCID(s):

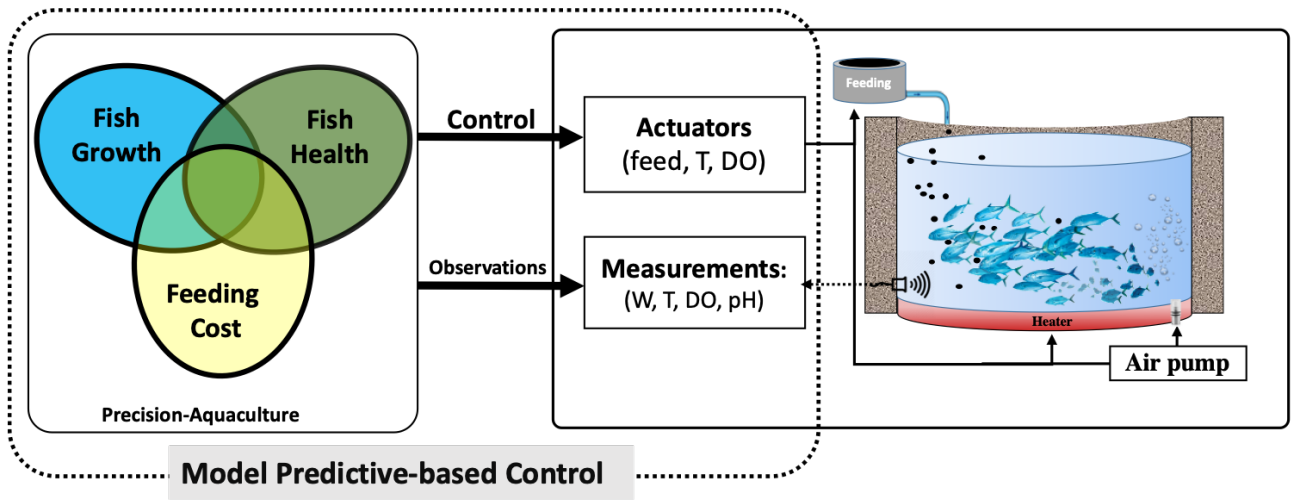

**Figure 1:** Precision aquaculture and optimal control framework.

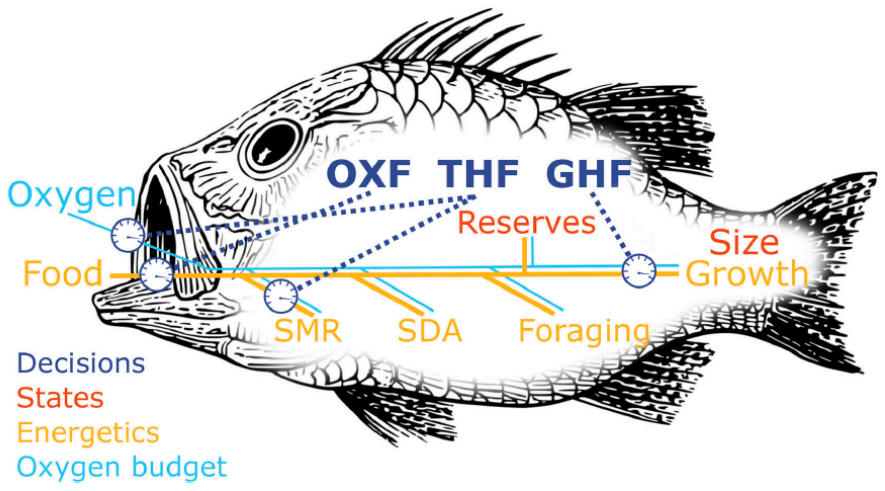

**Figure 2:** Dynamic energy budget model scheme in growing fish [25].

In this paper, model predictive control (MPC) strategies are investigated [14, 15, 13, 18, 19, 23, 10]. The main advantage of the MPC control is its ability to predict the behavior of controlled variables and take the right control action sequence that will optimize a predefined cost function over a prediction horizon. This cost function can optimize a quantity implying the state's dynamics and the constrained inputs, such as minimize the tracking error of a reference state. It can also optimize an external quantity such as the operating costs or any other measurable metric that is not directly related to the system dynamics. To the best of the authors' knowledge, the MPC based on the receding-horizon framework for reference growth trajectory has not been fully investigated in the aquaculture system.

The objective of this study is to provide a comparative assessment of three candidate MPC formulations to track a desired growth rate trajectory accounting for specific economic considerations and handling inputs constraints at each sampling time in an aquaculture environment. The optimal control problem studied aims at tracking a desired fish growth reference trajectory while penalizing the manipulated inputs in an aquaculture environment under a representative bioenergetic fish growth model. The first MPC formulation is based on receding-horizon that minimizes the tracking error while penalizing the feed ration, temperature, and dissolved oxygen; the second receding-horizon approach is formulated as the ratio between food quantity and fish weight gain the feed conversation ratio which defines the measure of the fish efficiency in converting feed mass into increased body mass, and the third MPC formulation maintains a tradeoff between reference growth trajectory tracking and dynamic energy and feeding pricing.

The outline of the rest of the paper is as follows. Section 2 describes the bioenergetic fish growth model of Nile tilapia (*Oreochromis niloticus*) incorporating the anabolism and catabolism growth coefficients. Section 3 formulates the MPC problem; the three candidate MPC formulations are presented. Section 4 presents the obtained results and discusses the findings for the three candidate MPC formulations, followed by a conclusion in Section 5.

## 2. Fish Growth Modeling

A representative two-term bioenergetic fish growth model that captures the dominant growth factors, including adequate fish size, feed ration and water temperature, is proposed in this work. The bioenergetic model is obtained from the dynamic energy budget. It presents a mechanistic basis for understanding an organism's energetics used to model the mass and energy flow through the fish from the uptake to usage for maintenance, reproduction, growth, and excretion [9, 11, 25], as illustrated in Fig 2. The model is expressed in terms of energy fluxes between the organism and the environment. It constitutes useful tools in the early stage of an aquaculture activity to carry the capacity of a system before installing new farms [24, 5] estimate production and feeding ration [2], or to optimize integrated multi-trophic aquaculture systems [20].

**Table 1**
Nomenclature and main parameters of the growth model

| Symbol | Description | Unit |
|---|---|---|
| $w$ | Fish weight | g |
| $t$ | Time | day |
| $m$ | Exponent of body weight for net anabolism | 0.67 |
| $n$ | Exponent of body weight for fasting catabolism | 0.81 |
| $f$ | Relative feeding rate | $0<f<1$ |
| $T$ | Temperature | $^0$C |
| $DO$ | Dissolved oxygen | mg/l |
| $UIA$ | Unionized ammonia | mg/l |
| $b$ | Efficiency of food assimilation | 0.62 |
| $a$ | Fraction of the food assimilated | 0.53 |
| $h$ | Coefficient of food consumption | $0.8\mathrm{g}^{1-\mathrm{m}}$/day |
| $k_{\mathbf{min}}$ | Coefficient of fasting catabolism | 0.00133N |
| $j$ | Coefficient of fasting catabolism | 0.0132N |
| $T_{opt}$ | Optimal average level of water temperature | $33^0$C |
| $T_{\mathbf{min}}$ | Minimum level of temperature | $24^0$C |
| $T_{\mathbf{max}}$ | Maximum level of temperature | $40^0$C |
| $UIA_{\mathbf{crit}}$ | Critical limit of UIA | 0.06mg/l |
| $UIA_{\mathbf{max}}$ | Maximum level of UIA | 1.4mg/l |
| $DO_{\mathbf{crit}}$ | Critical limit of DO | 0.3mg/l |
| $DO_{\mathbf{min}}$ | Minimum level of DO | 1mg/l |
| $r$ | Daily ration | g/day |
| $R$ | Maximal daily ration | 10% BWD |
| $BWD$ | Average body-weight per day | g/day |
| $\tau$ | Temperature factor | $0<\tau<1$ |
| $\sigma$ | Dissolved oxygen factor | $0<\sigma<1$ |
| $v$ | un-ionized ammonia factor | $0<v<1$ |
| $\rho$ | Photoperiod factor | $0<\rho<2$ |

According to Ursin's work [22], the fish growth model in both recirculating aquaculture systems and marine cages can be expressed as the difference between anabolism and catabolism [27, 26, 12, 7]. In this paper, a bioenergetic growth model is adopted for Nile tilapia cultured in fertilized marine ponds, incorporating available information in pond dynamic and fish physiology. The model includes the effects of different parameters such as water temperature, body size, un-ionized ammonia (UIA), dissolved oxygen (DO), photoperiod, and food availability [26]. Thus, the growth rate model of Nile tilapia is described as the difference between anabolism and catabolism [26]

$$\frac{\mathrm{d}w}{\mathrm{d}t} = \underbrace{\Psi\big(f,T,DO\big)v(UIA)\,w^m}_{\text{anabolism}} - \underbrace{k(T)}_{\text{catabolism}}\,w^n, \tag{1}$$

where $\Psi\big(f,T,DO\big)$ ($\mathrm{g}^{1-\mathrm{m}}\mathrm{day}^{-1}$) and $v(UIA)$ are the coefficients of anabolism and $k(T)$ ($\mathrm{g}^{1-\mathrm{n}}\mathrm{day}^{-1}$) is the coefficient of fasting catabolism expressed as

$$\Psi\big(f,T,DO\big) = h\rho f b(1-a)\tau(T)\sigma(DO), \quad \text{and} \quad k(T) = k_{\min}\exp\Big(j(T-T_{\min})\Big). \tag{2}$$

The effects of temperature $\tau(T)$, unionized ammonia $v(UIA)$ and dissolved oxygen $\sigma(DO)$ on food consumption are described, respectively [26].

$$\tau(T) = \begin{cases} \exp\left\{-\kappa\Big(\dfrac{T-T_{opt}}{T_{\max}-T_{opt}}\Big)^4\right\} & \text{if} \quad T > T_{opt}, \\ \exp\left\{-\kappa\Big(\dfrac{T_{opt}-T}{T_{opt}-T_{\min}}\Big)^4\right\} & \text{if} \quad T < T_{opt}, \end{cases}$$

where $\kappa = 4.6$.

$$v(UIA) = \begin{cases} 1 & \text{if} \quad UIA < UIA_{\text{crit}}, \\ \dfrac{UIA_{\text{max}} - UIA}{UIA_{\text{max}} - UIA_{\text{crit}}} & \text{if}\, UIA_{\text{crit}} < UIA < UIA_{\text{max}}, \\ 0 & \text{elsewhere.} \end{cases}$$

$$\sigma(DO) = \begin{cases} 1 & \text{if} \quad DO > DO_{\text{crit}}, \\ \dfrac{DO - DO_{\text{min}}}{DO_{\text{crit}} - DO_{\text{min}}} & \text{if} \quad DO_{\text{min}} < DO < DO_{\text{crit}}, \\ 0 & \text{elsewhere.} \end{cases}$$

Table 1 summarizes the nomenclature and the main parameters of the growth model [26].

The rate of growth (1) can be expressed in a compact form as follows

$$\frac{\mathrm{d}w}{\mathrm{d}t} = g\big(w, \underbrace{f, T, DO}_{u}, UIA\big), \tag{3}$$

where $w \in \mathbb{W} \subset \mathbb{R}$ denotes the state and $u = [u_1, u_2, u_3]^T$ is the input vector. $u \in \mathbb{U} \subset \mathbb{R}^3$ describes the manipulated control input vector of the fish growth model that depends on the feeding rate, temperature, and dissolved oxygen, respectively. The unionized ammonia function $v(UIA)$ is considered to be known and constant over the prediction horizon. The set of admissible input values $\mathbb{U}$ is compact. The relative feeding rate $f$ is formulated as the ratio between the daily ration $r$ and the maximal daily ration $R$ as follows

$$f = \frac{r}{R}.$$

The function $g : \mathbb{W} \times \mathbb{U} \to \mathbb{W}$ is locally Lipschitz on $\mathbb{W} \times \mathbb{U}$. The measurement growth state (3) is synchronously sampled at the current sampling time defined as $t_k = k\varepsilon$ where $k \in \mathbb{Z}^+$ is a positive integer and $\varepsilon > 0$ is the sampling period.

## 3. Optimization formulations

The bioenergetic fish growth model is highly nonlinear with multi-inputs, as defined in equation (3). The MPC provides the benefits of efficiently handling the multi-inputs of the growth model (3), considering all the constraints and nonlinearities and re-optimizing an $N$-step control sequence at each time step. All three MPC strategies studied in this work track the desired fish growth reference trajectory without compromising the energy costs. The optimal control problem can be formulated as a minimization with a finite-time prediction horizon as follows

$$\min_{u \in \mathcal{U}(\varepsilon)} J = \int_{t_k}^{t_{k+N}} \ell(\tilde{w}(\tau), w^d(\tau), u(\tau))\, \mathrm{d}\tau + \ell_T(\tilde{w}(t_{k+N}), w^d(t_{k+N}), N_o) \tag{4a}$$

$$\text{s.t} \quad \dot{\tilde{w}}(t) = g\big(\tilde{w}(t), u(t)\big) \tag{4b}$$

$$u(t) \in \mathbb{U}, \quad \forall t \in [t_k,\, t_{k+N}] \tag{4c}$$

$$\tilde{w}(t_k) = w(t_k), \quad \tilde{w}(0) = w(t_0) \tag{4d}$$

where $N$ is the prediction horizon of this MPC,
$\ell(\hat{w}(\tau), w^d(\tau), u(\tau))$ is the stage cost, $N_o$ is the horizon length of the terminal cost, $\ell_T(w, w^d, N_o)$ is the terminal cost. $\tilde{w}$ is the predicted state trajectory over the prediction horizon $[t_k,\, t_{k+N}]$ and $w^d$ is the desired reference live-weight growth trajectory. The potential growth rate profile $w^d$ is based on experimental data analysis and describes the rate achieved by a specific strain that satisfies all the nutritional requirements. $\mathcal{U}(\varepsilon)$ represents the set of piecewise constant functions described by the sampling period $\varepsilon$. The first control action $u(t_k)$ of the MPC (4) is implemented, and then the MPC horizon is rolled again over the next time step. Throughout the sampling period $[t_k,\, t_{k+N}]$, the first control action is applied in a sampled-and-hold fashion. The optimal solution to this optimization problem, which is defined for $[t_k,\, t_{k+N}]$ is denoted by $u_i^*(t\,|\,t_k), i = 1, \cdots, m$. The terminal cost $\ell_T(w, w^d, N_o)$ is constructed by calculating the predicted growth $\hat{w}(t\,|\,t_k)$ and input $\hat{u}(t\,|\,t_k)$ trajectories of system (3) with the initial condition $\hat{w}(t_k) = w(t_k)$. The predicted growth rate trajectory is obtained by integrating recursively (3) over the time interval $[t_k,\, t_{k+N_o}]$ for $l = 0, \cdots, N_{o-1}$ as follows

$$\dot{\hat{w}}(t|t_k) = g\big(\hat{w}(t|t_k), \hat{u}(t|t_k)\big) \tag{5a}$$

$$\hat{u}(t|t_k) \in \mathbb{U}, \quad \forall t \in [t_{k+l},\, t_{k+l+1}] \tag{5b}$$

$$\hat{w}(t|t_k) = w(t|t_k). \tag{5c}$$

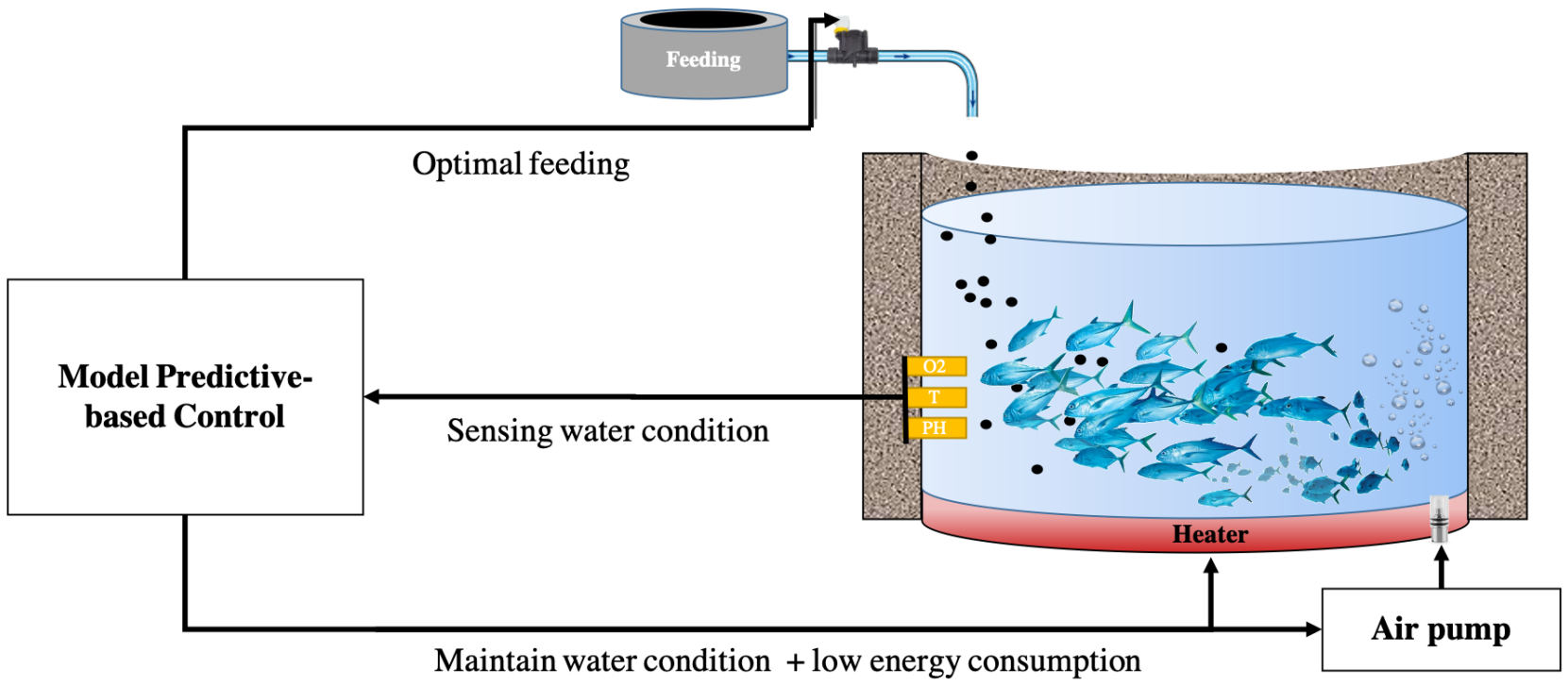

**Figure 3:** Model predictive control framework

The terminal cost $\ell_T(w, w^d, N_o)$ which is the cumulative performance over $N_o$ sampling periods is defined as follows

$$\ell_T(w(t_k), w^d(t_k), N_o) = \int_{t_k}^{t_{k+N_o}} \ell(\hat{w}(\tau), w^d(\tau), \hat{u}(\tau))\, \mathrm{d}\,\tau. \tag{6}$$

The three MPC optimization formulations are described as follows.

1. *Reference Trajectory Tracking, including Feeding and Energy Consumption Minimization* ($\mathrm{MPC}^1$): In this approach, the MPC strategy minimizes the growth rate tracking error while penalizing the food and energy quantities. We denote this cost function $J_{\mathrm{MPC}^1}$.
2. *Feed Conversion Ratio (FCR)* ($\mathrm{MPC}^2$): In this approach, the MPC optimization strategy minimizes the FCR as a standard metric to assess the aquaculture systems while penalizing the deviation of the actual growth state from the given reference growth trajectory. This metric is the ratio between food quantity and fish weight gain. This cost function is denoted $J_{\mathrm{MPC}^2}$.
3. *Reference Trajectory Tracking, including Economic Profitability Feeding, and Energy Consumption Minimization* ($\mathrm{MPC}^3$): In this approach, the MPC formulation tracks a given reference growth trajectory while reducing the economic profit and the total costs related to the feed and the electrical energy used for heating and oxygenation. This cost function is called $J_{\mathrm{MPC}^3}$.

The two first MPC formulations, which are the reference growth trajectory tracking including food, and energy consumption minimization $\mathrm{MPC}^1$ and the feed conversion ratio (FCR) $\mathrm{MPC}^2$ strategies consider objective functions that are not directly related to the economic cost.

### 3.1. Reference Trajectory Tracking, Food and Energy Consumption Minimization Strategy ($\mathrm{MPC}^1$)

The first MPC formulation tracks the desired fish growth trajectory while minimizing the feed ration, temperature, and dissolved oxygen. We formulate the MPC optimization problem for this strategy as follows

$$\min_{u \in \mathcal{U}(\varepsilon)} J_{\mathrm{MPC}^1} = \int_{t_k}^{t_{k+N}} \ell_1(\tilde{w}(\tau), w^d(\tau), u(\tau))\, \mathrm{d}\,\tau + \ell_T(\tilde{w}_{t_{k+N}}, w^d_{t_{k+N}}, N_o) \tag{7a}$$

$$\text{s.t} \quad \dot{\tilde{w}}(t) = g\big(\tilde{w}(t), u(t)\big) \tag{7b}$$
$$u(t) \in \mathbb{U}, \quad \forall t \in [t_k,\ t_{k+N}] \tag{7c}$$
$$\tilde{w}(t_k) = w(t_k) \tag{7d}$$

where the stage cost $\ell_1$ is defined as follows

$$\ell_1(\tilde{w}(\tau), w^d(\tau), u(\tau)) = \left\| \frac{\tilde{w}(\tau) - w^d(\tau)}{w^d(\tau)} \right\|^2 + \lambda \left\| u(\tau) \right\|^2, \tag{8}$$

and $\lambda$ is a positive regularization term to assess the control inputs preference. $\lambda$ is tuned empirically such that a good compromise between tracking error performance and fast tracking response is achieved over the entire prediction horizon.

## 3.2. Feed Conversion Ratio (FCR) Strategy (MPC$^2$)

Fish feeding is an essential component of fish farming; the reduction of the FCR helps in more efficient use of fishmeal, energy consumption, and fish oil, which has primarily been achieved through improved management. In this approach, the MPC formulation optimizes the feed conversion ratio (FCR), which is the mass of the food eaten divided by the mass body gain while tracking the desired growth reference trajectory. The FCR describes the quantity of feed used to the fish organisms under satisfactory conditions for its development. The cost function is defined as follows

$$\min_{u \in \mathcal{U}(\epsilon)} J_{\text{MPC}^2} = \int_{t_k}^{t_{k+N}} \ell_2(\tilde{w}(\tau), w^d(\tau), u(\tau))\, \mathrm{d}\tau + \ell_T(\tilde{w}_{t_{k+N}}, w^d_{t_{k+N}}, N_o) \tag{9a}$$
$$\text{s.t} \quad \dot{\tilde{w}}(t) = g\big(\tilde{w}(t), u(t)\big) \tag{9b}$$
$$u(t) \in \mathbb{U}, \quad \forall t \in [t_k,\ t_{k+N}] \tag{9c}$$
$$\tilde{w}(t_k) = w(t_k) \tag{9d}$$

where $\ell_2$ represents the FCR defined as

$$\ell_2(\tilde{w}(\tau), w^d(\tau), u(\tau)) = \frac{u_1(\tau)}{\Delta w(\tau)}, \tag{10}$$

with $u_1$ is the relative feeding rate, and $\Delta w(\tau)$ is the net weight gain with respect to the initial and final weights. $\Delta w(\tau)$ is sampled synchronously at time instants $t_k$ over the entire prediction horizon. FCR formulated as an optimal feeding strategy provides a good indicator of farming efficiency, economic and environmental performance since this index successfully minimizes the deviation of the growth rate performance and the use of feeding resources supplied.

## 3.3. Reference Trajectory Tracking, Economic Profitability Food and Energy Consumption Minimization Strategy (MPC$^3$)

The third MPC formulation minimizes the economic profit and the energy consumption costs, including the feeding, heating, and oxygenation of the aquaculture environment system. The MPC optimization problem for this strategy is defined as follows

$$\min_{u \in \mathcal{U}(\epsilon)} J_{\text{MPC}^3} = \int_{t_k}^{t_{k+N}} \ell_3(\tilde{w}(\tau), w^d(\tau), u(\tau))\, \mathrm{d}\tau + \ell_T(\tilde{w}_{t_{k+N}}, w^d_{t_{k+N}}, N_o) \tag{11a}$$
$$\text{s.t} \quad \dot{\tilde{w}}(t) = g\big(\tilde{w}(t), u(t)\big) \tag{11b}$$
$$u(t) \in \mathbb{U}, \quad \forall t \in [t_k,\ t_{k+N}] \tag{11c}$$
$$\tilde{w}(t_k) = w(t_k), \tag{11d}$$

where $\ell_3$ represents the stage cost associated with an economic profitability term defined as follows

$$\begin{aligned}\ell_3(\tilde{w}(\tau), w^d(\tau), u(\tau)) = {} & B_1\big(w(\tau) - w^d(\tau)\big)^2 \quad \text{tracking error cost} \\ & + B_2 u_1(\tau)^2 \quad \text{feeding cost} \\ & + B_3 u_2(\tau)^2 \quad \text{heating cost} \\ & + B_4 u_3(\tau)^2 \quad \text{oxygenation cost}\end{aligned}$$

**Table 2**
Parameters value of the cost functions

| Parameters | Value |
|---|---|
| $\lambda$ | 0.1 |
| $\alpha$ | 100 |
| $P_s$ | 1.2 \$/kg |
| $P_f$ | 0.4\$/kg |
| $R$ | 10% |
| $\beta_1 = \beta_2$ | 0.1 |
| $P_e$ | 0.14\$/kWh |
| $c_p$ | 4.2J/kgC |
| $L$ | 4541 |
| $m$ | 1 |
| $P_{\text{max}}$ | 0.102kWh |

For this comparative study, the weight $B_1$ is considered constant over the prediction horizon. The cost weights $B_2$, $B_3$, and $B_4$ vary with the time and account for the price of the feeding, heating, and oxygenation resources.

The cost weights $B_1$, $B_2$, $B_3$ and $B_4$ of this third MPC optimization strategy are explicitly defined as follows

$$\ell_3(\tilde{w}(\tau), w^d(\tau), u(\tau)) = \alpha\Big(P_s(w(\tau) - w^d(\tau)\Big)^2 + \big(P_f R u_1(\tau)\big)^2 + \beta_1\Big(\frac{P_e c_p L m \Delta u_2(\tau)}{3600}\Big)^2 + \beta_2\Big(24 P_e P_{\max} u_3(\tau)\Big)^2, \quad (12)$$

where $\alpha$ is a regularization term to improve growth tracking error performance. $P_s$ is the fish selling price per kg [1]. $P_f$ is the fish food price per kg, $u_1 = f$ represents the feeding rate, $R$ is the maximal daily ration. $\beta_1$ and $\beta_2$ are regularization terms defining the heater's daily operation duration ratio and the air pump, respectively. $P_e$ is the electricity price per kWh, $c_p$ is the specific heat of the tank water, $L$ is the tank volume in liters, $m = 1$ is the water mass, and $\Delta u_2 = \Delta T$ represents the temperature difference in $^{o}$C [16]. $P_{\max}$ is the maximal electrical power of the air pump, and $u_3 = DO$ represents the dissolved oxygen level.

## 4. Numerical Simulations

This section presents a comparative analysis of the three proposed candidate MPC formulations using the fish growth model and interprets the obtained results. The parameters of the Nile tilapia growth model are set based on the values provided in [26]. Besides, the potential growth reference tracking profile $w^d$ is based on experimental data analysis and describes the rate achieved by a specific strain that satisfies all the nutritional requirements [3]. The three MPC formulations are implemented using the Open Optimal Control Library [8]. The values of the parameters used in the numerical results are summarized in Table 2.

To compare the three MPC optimization strategies' performance, we consider the feed conversion ratio, profit, and profit percentage as performance evaluation metrics to assess the aquaculture systems. These metrics are defined as follows

$$\textbf{FCR} = \frac{\text{total feed quantity (kg)}}{\text{final weight (kg)} - \text{initial weight (kg)}}, \qquad \textbf{Profit} = \text{revenue} - \text{total costs}, \quad (13)$$

and

$$\textbf{Profit percentage} = \frac{\text{revenue}}{\text{total costs}}. \quad (14)$$

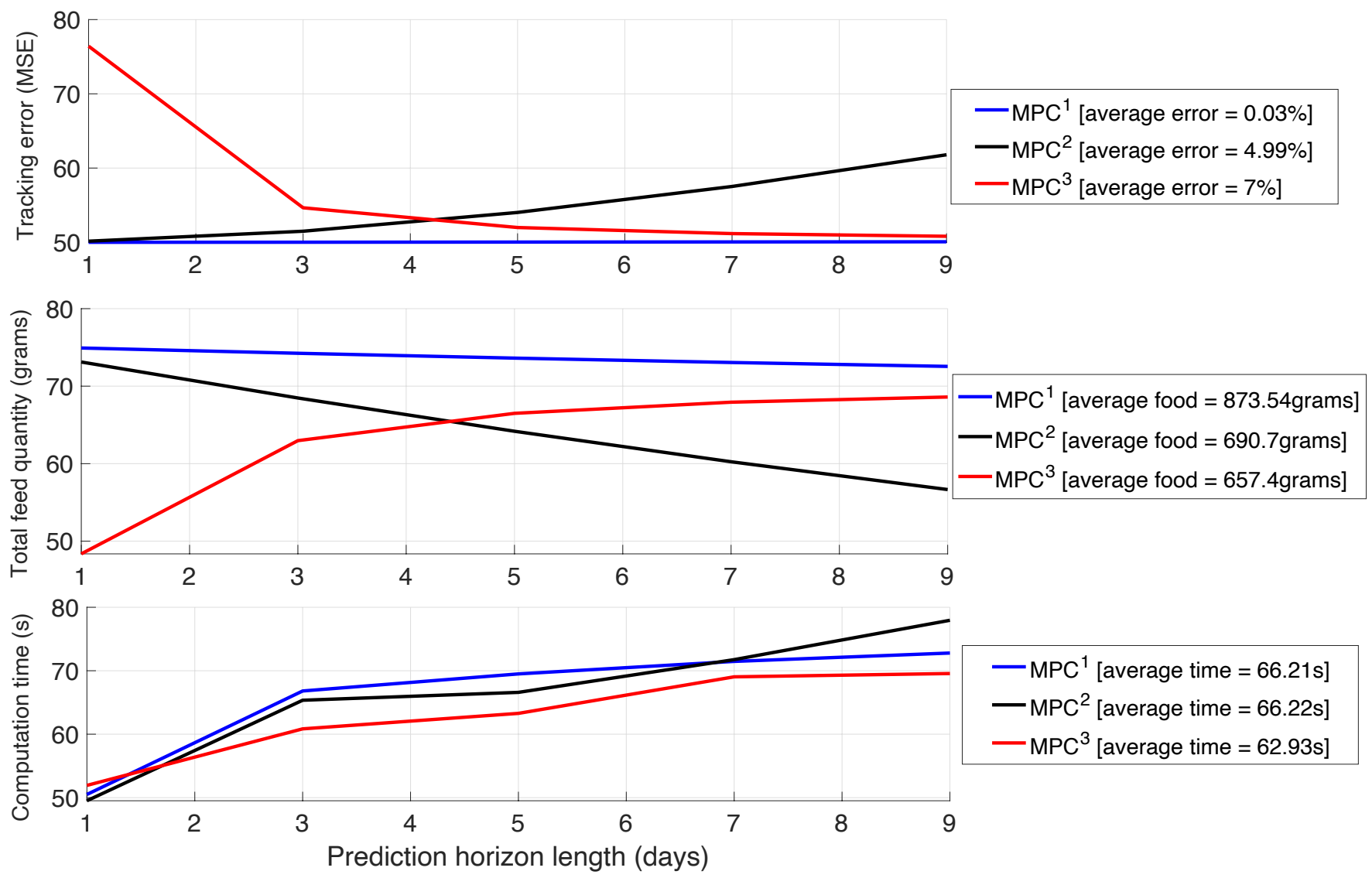

**Figure 4:** Effect of the horizon length on the three MPC formulations.

**Table 3**
Performance Comparison of the Three Candidate MPC Formulations with and without Measurement Noise.

| Horizon Length | Noise | Controller | Tracking error (MSE) | Number of fish | Fish Weight(g) | Feed Quatity (g) | Elapsed Time (s) | Revenue (USD) | Feed Cost (USD) | Heating Cost (USD) | Oxygenation Cost (USD) | Profit (USD) | Profit percentage | FCR |
|---|---|---|---|---|---|---|---|---|---|---|---|---|---|---|
| 3 days | No | MPC$^2$ | 1.49 | 1000 | 386.18 | 768.29 | 51.41 | 463.42 | 307.32 | 11.06 | 3.67 | 141.37 | 43.89 | 1.53 |
| | | MPC$^1$ | **0.01** | 1000 | **427.61** | 883.34 | 56.79 | 513.13 | 353.34 | 11.25 | 3.67 | 144.87 | 39.34 | 1.44 |
| | | MPC$^3$ | 4.63 | 1000 | 343.77 | 658.77 | 55.83 | 412.53 | 263.51 | 11.32 | 3.38 | **134.31** | **48.27** | **1.64** |
| | Yes (50dB) | MPC$^2$ | 0.21 | 1000 | 417.47 | 719.17 | 71.29 | 500.97 | 287.67 | 14.27 | 3.85 | 195.18 | 63.83 | 1.29 |
| | | MPC$^1$ | **0.01** | 1000 | **427.61** | 935.79 | 63.05 | 513.14 | 374.32 | 22.43 | 3.67 | 112.72 | 28.15 | 1.47 |
| | | MPC$^3$ | 4.82 | 1000 | 344.93 | 654.97 | 57.84 | 413.91 | 261.99 | 11.30 | 3.38 | **137.24** | **49.60** | **1.62** |

## 4.1. Effect of the Horizon Length on the Performance and Computation Time of the MPC Formulations

The prediction horizon plays an essential role in the optimization problem of the three MPC formulations. Fig. 4 shows the tracking error performance improves with the increase of the prediction horizon for the third MPC strategy. However, the feeding quantity improves with the rise of the prediction horizon for the first MPC strategy. Subsequently, the computation time increases with the length of the prediction horizon for all MPC strategies. Overall, the third MPC strategy achieves the best average performance for the tracking error performance and computational cost. Consequently, the prediction horizon length of $N = N_o$ =3days is used for all the following simulations for this comparative study.

## 4.2. Effect of the Measurement Noise on the Performance of the MPC Formulations

In real scenarios, the feeding and heating systems are not accurate enough, which can be interpreted as Gaussian noise disturbances in the feeding quantity and temperature in tanks. We simulate this performance by adding zero-mean Gaussian noise to the feeding and temperature control actuators over each integration step. Table 3 shows that the first MPC controller gives the best tracking performance but with higher feeding quantity than the second and third MPC strategies. However, the third MPC optimization achieves the best economic cost-benefit ratio, as illustrated by the highest profit percentage. Figs. 5, 6, and 7 show that the first MPC strategy provides the best tracking of a given fish growth, which might represent a healthy growth profile to track, specifically in the early stage of the fish age where the mortality ratio is high. The third MPC optimization can be used for commercial uses, where the final profit is a significant concern. Besides, the third MPC strategy can help manage other costs streams as long as they are measurable. Finally, the second MPC strategy seems to have the best compromise between the two abilities as it can achieve a good tracking performance with an acceptable economic profit. Moreover, the second MPC strategy reflects an existing metric in the aquaculture sector as a metric, which can be interpreted easily.

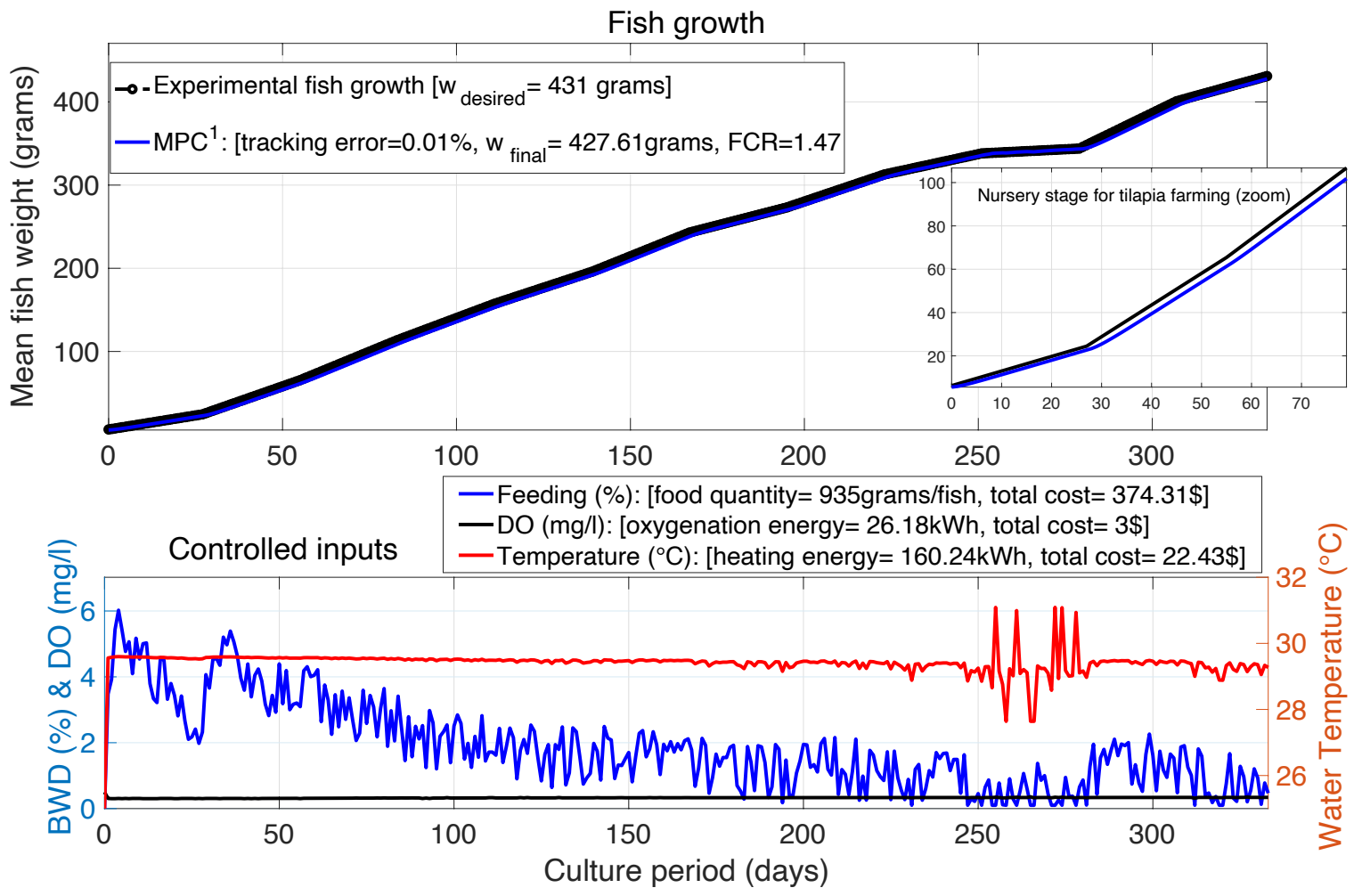

**Figure 5:** Fish growth trajectory with minimum food and energy consumption: MPC[1].

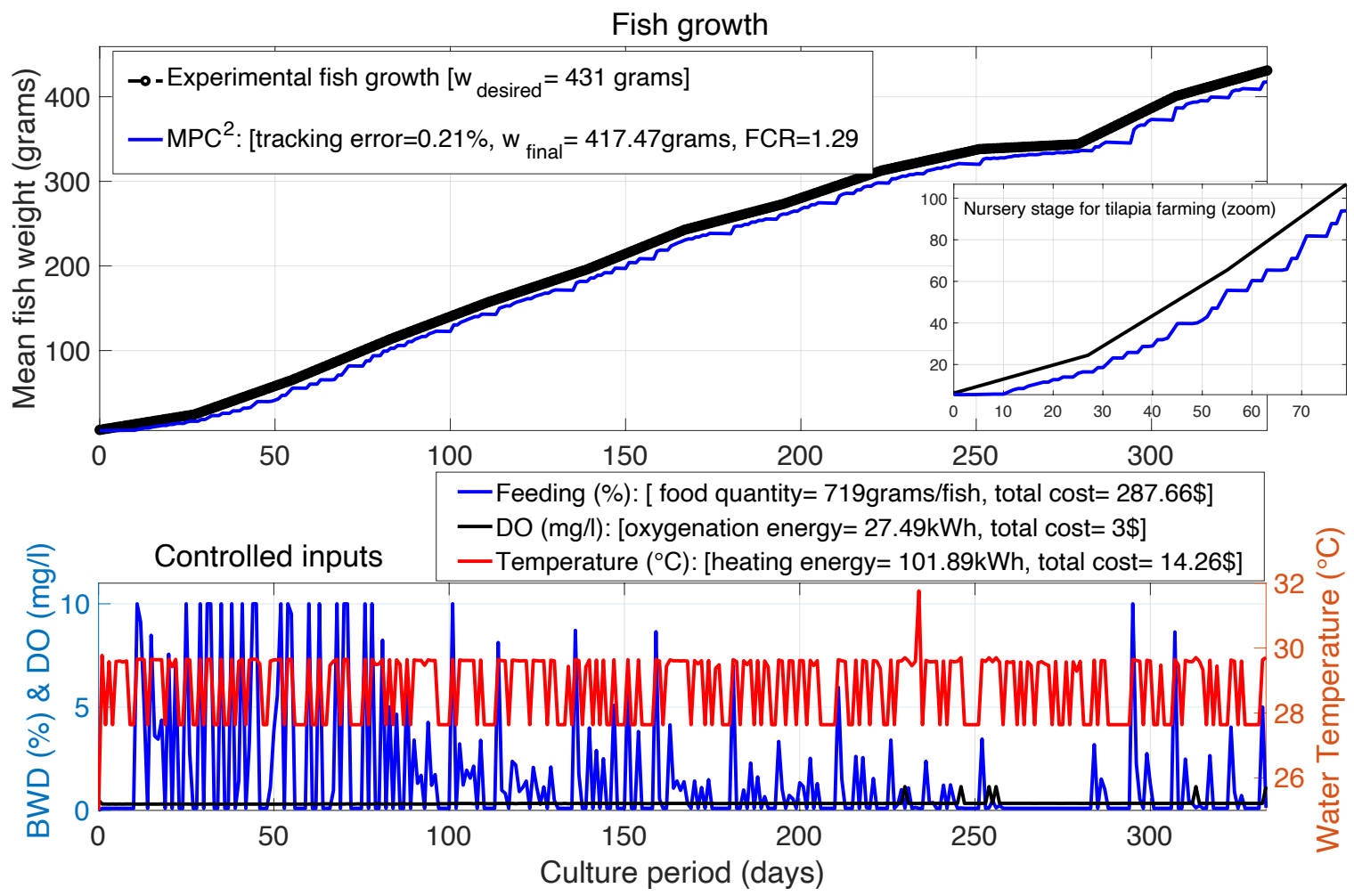

**Figure 6:** Fish growth trajectory with minimum feed conversion ratio: MPC[2].

## 5. Conclusion

All three proposed MPC strategies provided useful driving growth trajectories, resulting in improved economic profit and energy consumption. The obtained results show that the proposed MPC strategies can successfully meet the target reference growth trajectory, which allows the fish farmer to choose the most suitable method as a baseline control strategy for his own needs. For instance, the first MPC approach presents the best tracking ability, but it consumes a higher feed quantity than the two other MPC strategies. The third MPC strategy shows the best economic profit with lower tracking ability. However, the second MPC strategy achieves the best compromise between good tracking and high profitability. The implementation of the proposed MPC controllers is available online and downloadable from: https://github.com/EMANG-KAUST/Economic-Model-predictive-control-for-Aquacuture.git

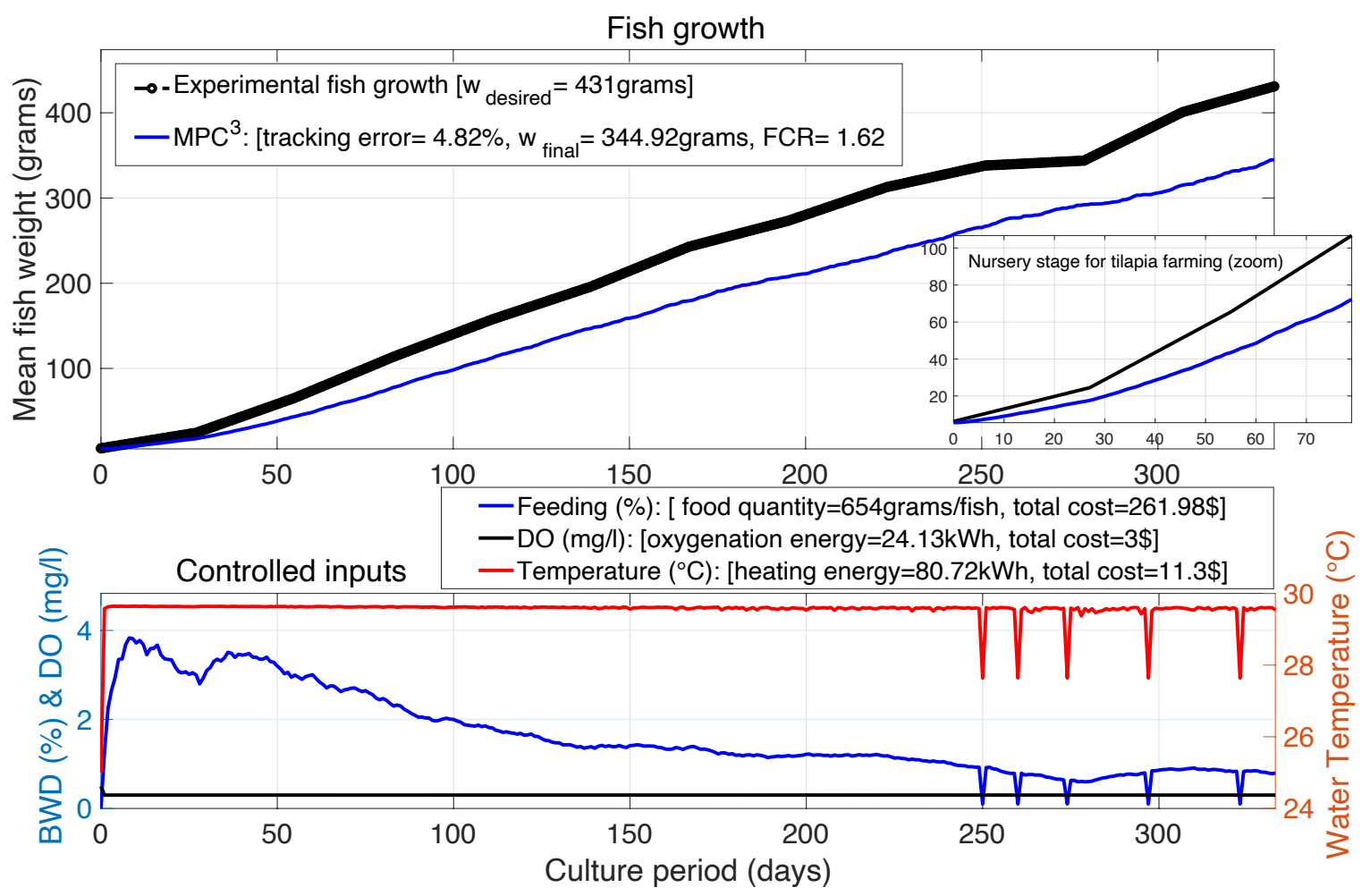

**Figure 7:** Fish growth trajectory including economic profitability and minimum food and energy consumption: MPC$^3$.